\documentclass[pra,floatfix,twocolumn,superscriptaddress]{revtex4}

\usepackage{amsmath,amssymb,amsfonts,amsthm}
\usepackage{graphicx} 
\usepackage{dcolumn} 
\usepackage{bm} 
\usepackage{color}
\usepackage{hyperref}

\usepackage{bbm}
\usepackage{exscale}
\usepackage{epstopdf}

\newcommand{\beq}{\begin{equation}} 
\newcommand{\eeq}{\end{equation}}
\newcommand{\bqa}{\begin{eqnarray}} 
\newcommand{\eqa}{\end{eqnarray}}
\newcommand{\nn}{\nonumber}

\newcommand{\dg}{^\dagger}
\newcommand{\rt}[1]{\sqrt{#1}\,}

\newcommand{\bra}[1]{\langle{#1}|} 
\newcommand{\ket}[1]{|{#1}\rangle}
\newcommand{\an}[1]{\langle{#1}\rangle}
\newcommand{\ban}[1]{\big\langle{#1}\big\rangle}

\newcommand{\op}[2]{\left|{#1}\rangle \langle{#2}\right|}

\newcommand{\sch}{Schr\"odinger} 
\newcommand{\hei}{Heisenberg}

\newcommand{\ito}{It\^o} 
\newcommand{\str}{Stratonovich}

\newcommand{\Tr}{{\rm Tr}}

\newcommand{\piez}{\hat{\pi}^z}
\newcommand{\pieplus}{\hat{\pi}^+}
\newcommand{\piemin}{\hat{\pi}^-}

\newcommand{\ann}{\hat{a}}
\newcommand{\adg}{\hat{a}\dg}

\newcommand{\annt}{\hat{a}^2}
\newcommand{\adgt}{\hat{a}\dg{}^2}

\newcommand{\upp}{\Uparrow}
\newcommand{\ddn}{\Downarrow}

\newcommand{\Hhat}{\hat{H}}

\newcommand{\Vhat}{\hat{V}}

\newcommand{\ddt}{\frac{d}{dt}}

\newcommand{\Pip}{\tilde{\Pi}}

\newcommand{\kB}{k_{\text{\tiny B}}}
\newcommand{\Dcal}{{\cal D}}
\newcommand{\Lcal}{{\cal L}}

\newcommand{\wo}{\omega_0}

\newcommand{\Pihat}{\hat{\Pi}}

\newcommand{\abar}{\bar{a}}

\newcommand{\Acal}{{\cal A}}
\newcommand{\Phase}{{\cal P}_\varphi}

\newcommand{\Ecal}{{\cal E}}

\newcommand{\kup}{\kappa_{\uparrow}}
\newcommand{\kdn}{\kappa_{\downarrow}}
\newcommand{\kupp}{\kappa_\Uparrow}
\newcommand{\kddn}{\kappa_\Downarrow}

\newcommand{\nhat}{\hat{n}}

\newcommand{\rhoin}{\rho_{\rm in}}
\newcommand{\rhoout}{\rho_{\rm out}}

\begin{document}

\title{Phase-preserving linear amplifiers not simulable by the parametric amplifier}

\author{A. Chia}
\affiliation{Centre for Quantum Technologies, National University of Singapore}

\author{M. Hajdu\v{s}ek}
\affiliation{Keio University Shonan Fujisawa Campus, Kanagawa, Japan}

\author{R. Nair}
\affiliation{School of Physical and Mathematical Sciences, Nanyang Technological University, Singapore}
\affiliation{Complexity Institute, Nanyang Technological University, Singapore}

\author{R. Fazio}
\affiliation{Abdus Salam ICTP, Strada Costiera, Trieste, Italy.}

\author{L. C. Kwek}
\affiliation{Centre for Quantum Technologies, National University of Singapore}
\affiliation{National Institute of Education, Nanyang Technological University, Singapore}

\author{V. Vedral}
\affiliation{Centre for Quantum Technologies, National University of Singapore}
\affiliation{Department of Physics, University of Oxford, UK}

\date{\today}

\begin{abstract}

It is commonly accepted that a parametric amplifier can simulate a phase-preserving linear amplifier regardless of how the latter is realised [C.~M.~Caves~\emph{et al.}, Phys.~Rev.~A~{\bf 86}, 063802 (2012)]. If true, this reduces all phase-preserving linear amplifiers to a single familiar model. Here we disprove this claim by constructing two counterexamples. A detailed discussion of the physics of our counterexamples is provided. It is shown that a \hei-picture analysis facilitates a microscopic explanation of the physics. This also resolves a question about the nature of amplifier-added noise in degenerate two-photon amplification.

\end{abstract}


\maketitle

\emph{Introduction.---\,}Linear amplification has long been an integral part of quantum measurements whereby a weak signal is amplified to a detectable level \cite{CTD+80,CDG+10}. Due to advances in quantum optics and quantum information, linear amplifiers are now also seen as a facilitating component of many useful tasks such as state discrimination \cite{ZFB10}, quantum feedback \cite{VMS+12}, metrology \cite{HKL+14}, and entanglement distillation \cite{RL09,XRLWP10}. New paradigms of amplification such as heralded probabilistic amplification \cite{RL09,ZFB10,CWA+14,HZD+16} and photon number amplification \cite{PvE19} are being actively researched for these and other applications.

Much attention has been given to the application and construction of linear amplifiers \cite{CTD+80,CDG+10}, and their fundamental quantum noise limits have been known for a long time \cite{Cav82maintext}. A relatively recent foundational development, however, is the claim that a parametric amplifier can simulate any phase-preserving linear amplifier regardless of how it is realised \cite{CCJP12}. This statement is significant as it replaces the set of all phase-preserving linear amplifiers by a single familiar model. Either proving it or falsifying it is thus of fundamental importance to our understanding of deterministic amplifiers.  It would also clarify the status of the parametric amplifier (henceforth abbreviated as paramp). More specifically, is it possible to find phase-preserving linear amplifiers which cannot be simulated by the paramp? If so, what differentiates such amplifiers from those that can be simulated by the paramp? 
%
%
%
%
%
%

\begin{figure}[t]
\centerline{\includegraphics[width=0.45\textwidth]{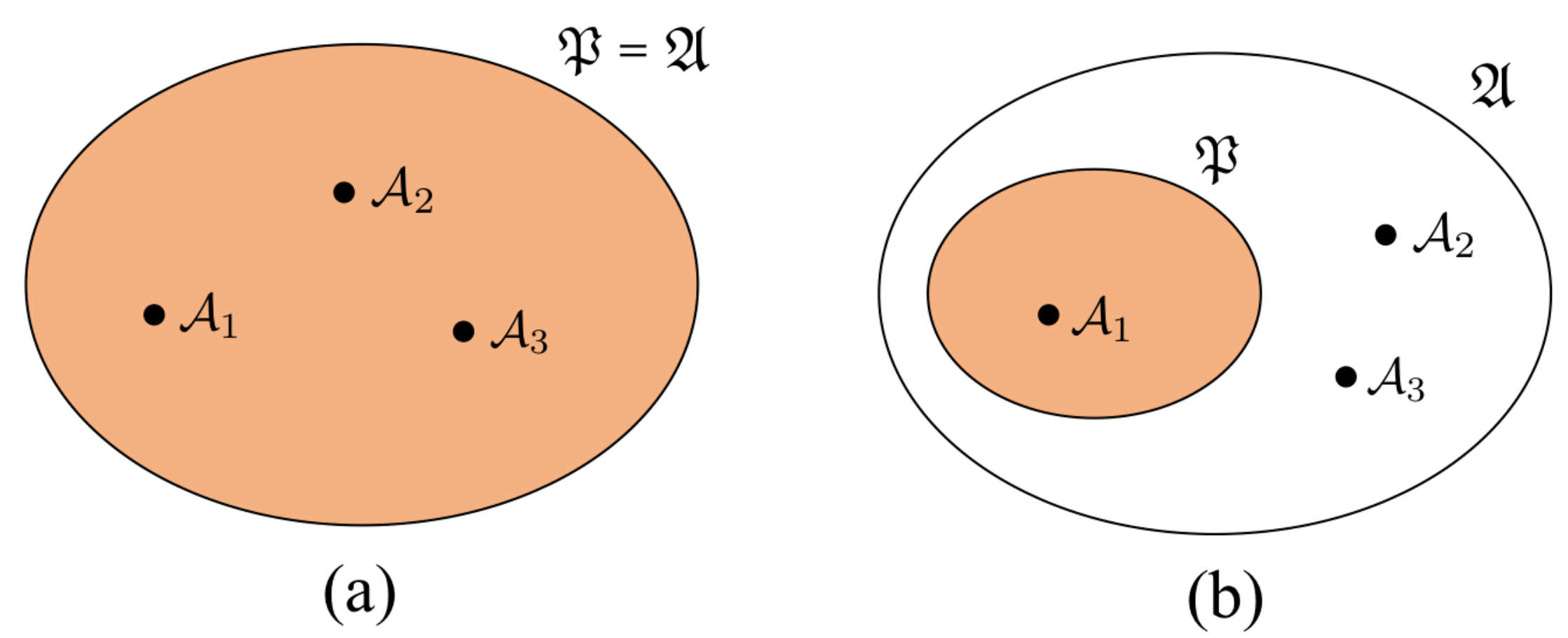}}
\caption{\label{MainResult} $\frak{A}$ denotes the set of all phase-preserving linear amplifiers [defined by (i)--(iii)] while those that are paramp simulable are in $\frak{P}$ [defined by \eqref{ParampConj} and coloured in orange]. The amplifiers $\Acal_1$, $\Acal_2$, and $\Acal_3$ are defined by generators given in \eqref{Example1}, \eqref{GeneratorForNIA}, and \eqref{ThreePhotonGenerator} respectively. (a) Accepted status of the paramp: $\frak{A}=\frak{P}$ \cite{CCJP12}. (b) Result of this paper: $\frak{P} \subsetneq \frak{A}$.}
\end{figure}

In this work, we provide answers to these questions. We provide as counterexamples two families of physically-realisable linear amplifiers which are phase preserving but cannot be simulated by the paramp. The inner workings of such amplifiers are then studied, revealing that the physical mechanism of multiplicative noise leads to amplifiers that are not simulable by the paramp. This delineates the boundary and status of the paramp in linear-amplifier theory. Our main result is summarised in Fig.~\ref{MainResult}. As a corollary, we also gain understanding on the nature of noise in nonlinear amplifiers. 

\emph{Definitions.---\,} We begin by making the above statements precise. We specify an amplifier by a map $\Acal$ which transforms the state of an input signal $\rhoin$ to a new state at its output $\rhoout=\Acal\,\rhoin$. Throughout this paper the signal itself will be represented by the single-mode bosonic annihilation operator $\ann$ acting on Hilbert space $\mathbbm{H}_A$. An amplifier is said to be 
\begin{itemize}
	\item[(i)] \emph{Physical} if $\Acal$ is completely positive and trace preserving.
	\item[(ii)] \emph{Linear} if $\Acal$ is such that $\an{\ann}_{\rm out}\equiv\Tr[\ann\Acal\rhoin]=g\an{\ann}_{\rm in}$ for all  $\rhoin$.
	\item[(iii)] \emph{Phase preserving} if the gain $g$ is real valued.
\end{itemize}
We denote the set of amplifiers satisfying (i)--(iii) by $\frak{A}$ \footnote{An additional requirement -- which we call \emph{phase covariance} \cite{CHN+19} -- has been mentioned in Sec.~III of \cite{CCJP12}, but both the universality claim and purported proof of it do not impose this requirement. In any case, both the counterexamples to the paramp conjecture that we present satisfy this property as well \cite{CHN+19}.}. A member of $\frak{A}$ is given by $\Acal_1(t)=\exp(\Lcal_1 t)$, where
\begin{align}
\label{Example1}
	\Lcal_1 = \kup \, \Dcal[\adg] + \kdn \, \Dcal[\ann]  \; ;  \quad \; \kup > \kdn \ge 0  \; ,
\end{align}
and $\Dcal[\hat{A}]\hat{B}\equiv\hat{A}\hat{B}\hat{A}\dg-\hat{A}\dg\hat{A}\hat{B}/2-\hat{B}\hat{A}\dg\hat{A}/2$. By virtue of its Lindblad form, \eqref{Example1} generates a family of completely-positive trace-preserving maps $\{\Acal_1(t)\}_t$ for fixed $\kup$ and $\kdn$ \cite{Lin76}. This is the familar master equation model of a linear amplifier \cite{BP02,Car02,DH14,Aga13,SZ97}.  It is not too difficult to show that $\Acal_1(t)$ is linear and phase preserving for any $t$ \cite{Aga13}.

\emph{Parametric amplifier.---\,}The paramp is a device with an internal degree of freedom represented by the bosonic annihilation operator $\hat{b}$ acting on $\mathbbm{H}_B$. Its inital state is denoted by $\sigma$. The paramp map $\Ecal$ is defined via the two-mode squeeze operator $\hat{S}=\exp[\,r(\ann\,\hat{b}-\adg\hat{b}\dg)]$ as
\begin{align}
\label{ParampMap}
	\rhoout = \Ecal \, \rhoin = \Tr_B\big[ \hat{S}\, \rhoin \otimes \sigma \,\hat{S}\dg \big]  \; , 
\end{align}
where $\Tr_{\rm B}$ denotes a partial trace over $\mathbbm{H}_B$. The gain of the paramp may be shown to be $G=\cosh r$ where $r$ is the squeezing parameter \cite{CCJP12}. This finally brings us to the universality claim of the paramp \cite{CCJP12}: Given any physical linear phase-preserving amplifier $\Acal$,  one can always find a $\sigma$ and $G$ of the paramp such that its output state is identical to the output state from $\Acal$ for  any input $\rhoin$, i.e., 
\begin{align}
\label{ParampConj}
	\exists \; \sigma, \, G\!: \,  \Ecal=\Acal \; , \quad  \forall \, \Acal \in \frak{A}.
\end{align}
If we denote the set of amplifiers that are paramp simulable by $\frak{P}$, \eqref{ParampConj} states that $\frak{P}=\frak{A}$ [shown in Fig.~\ref{MainResult}(a)].

\emph{Counterexamples.---\,}We consider first the family of maps $\Acal_2(t)=\exp(\Lcal_2\, t)$ generated by 
\begin{align}
\label{GeneratorForNIA}
	\Lcal_2 = \frac{\gamma}{2} \; \big( \, \Dcal\big[\ann^2\big] + \Dcal\big[\adg{}^2\big] \, \big); \;  \; \gamma > 0.
\end{align}
By virtue of its Lindblad form, $\{\Acal_2(t)\}_t$ is a physically valid family of maps for a fixed $\gamma$. Consider a particular member of this family  $\Acal_2=\exp(\Lcal_2\, t_0)$ for some choice of $t_0$. A straightforward calculation shows that this produces a linear amplifier $\an{\ann}_{\rm out}=g\,\an{\ann}_{\rm in}$ where $g=\exp(\gamma t_0)$. This establishes that $\Acal_2 \in \frak{A}$.

For the paramp $\Ecal$ to be equivalent to $\Acal_2$, it is necessary that the moments of $\ann$ at the output from both amplifiers be identical for an arbitrary input state $\rhoin$. Here we show that this cannot be satisfied by considering the output amplitude and photon-number moments corresponding to $\Ecal$ and $\Acal_2$. For $\Acal_2$ they are \cite{CHN+19}: 
%
\begin{align} \label{<a>NoiseInduced}
	\an{\ann}_{\rm out} = {}& g\,\an{\ann}_{\rm in}  \; ,  \\
\label{<n>NoiseInduced}
	\ban{\hat{n}}_{\rm out} = {}& g^4 \, \ban{\hat{n}}_{\rm in} + \frac{g^4-1}{2}  \; ,
\end{align}
where $\hat{n}=\adg\ann$. The same quantities for the paramp are \cite{CCJP12}:
\begin{align}
	\an{\ann}_{\rm out} = {}& G\,\an{\ann}_{\rm in} + \rt{G^2-1}\,\an{\hat{b}} \; ,   \\
\label{<n>Parametric}
	\ban{\hat{n}}_{\rm out} = {}& G^2 \ban{\hat{n}}_{\rm in} + \big(G^2-1\big)\,\ban{\hat{b}\,\hat{b}\dg}  \; ,
\end{align}
where all moments involving $\hat{b}$ are taken with respect to its internal state $\sigma$ while those involving $\ann$ are taken with respect to $\rhoin$. To ensure that the two amplifiers give identical $\an{\ann}_{\rm out}$ for any $\rhoin$ we must choose $\ban{\hat{b}}=0$ and set $G=g$. Now consider an input signal prepared in some state, say $\rho_1$, with average photon number $\an{\hat{n}}_1$. It is necessary that $\Acal_2$ and $\Ecal$ output the same photon number when applied to $\rho_1$, i.e.
\begin{align}
\label{<n(t)>1}
 	g^4 \ban{\hat{n}}_1 + \frac{g^4-1}{2} = {}& g^2 \ban{\hat{n}}_1 + (g^2-1) \, \ban{\hat{b}\,\hat{b}\dg}   \;.
\end{align}
Similarly we may consider another input state $\rho_2$ with a different average photon number $\an{\hat{n}}_2$. The same requirement leads to
\begin{align} 
\label{<n(t)>2}
	 g^4 \ban{\hat{n}}_2 + \frac{g^4-1}{2} = {}& g^2 \ban{\hat{n}}_2 + (g^2-1) \, \ban{\hat{b}\,\hat{b}\dg}   \;.
\end{align}
Subtracting \eqref{<n(t)>2} from \eqref{<n(t)>1} we get
\begin{align}
\label{<n>1-<n>2}
	g^4 \Big[ \ban{\hat{n}}_1 - \ban{\hat{n}}_2 \Big] = g^2 \Big[ \ban{\hat{n}}_1 - \ban{\hat{n}}_2 \Big]  \;.
\end{align}
Equation \eqref{<n>1-<n>2} clearly cannot be satisfied unless $g=1=G$ (which means no amplification). Thus, the paramp cannot be a universal model for $\frak{A}$. Note that it is the difference in how $\an{\hat{n}}_{\rm out}$ scales with $g$ in the two types of amplifiers that makes $\Ecal \ne \Acal_2$. To the best of our knowledge, this is the first time that a phase-preserving linear amplifier has been shown to fall outside the reach of the paramp.

It is natural to wonder whether the family of amplifiers  $\{\Acal_2(t)\}_t$ is something of a special case. Another family of counterexamples $\{\Acal_3(t)\}_t$ with $\Acal_3(t) = \exp(\Lcal_3 t)$ is derived from the generator
\begin{align}
\label{ThreePhotonGenerator}
	\Lcal_3 = \frac{\gamma}{9} \; \big( \, \Dcal\big[\ann^3\big] + \Dcal\big[\adg{}^3\big] \, \big) + \gamma \; \Dcal\big[\ann^2\big]  \;,  \quad   \gamma > 0  \; .
\end{align}
Physical realizability follows immediately from the Lindblad form of \eqref{ThreePhotonGenerator}, while properties (ii) and (iii) are shown in Ref.~\cite{CHN+19}. We have chosen the coefficients in \eqref{ThreePhotonGenerator} so that $\Acal_3(t)$ has the same gain $g = \exp(\gamma t)$ as $\Acal_2(t)$. In this case a simple analytic expression like \eqref{<n>NoiseInduced} cannot be found for its average output photon number. It is nevertheless possible to show that $\Acal_3(t)$ leads to an average output photon number which is irreproducible by the paramp \cite{CHN+19}.

\emph{Physical properties---\,}We now turn to the question of what differentiates amplifiers which are paramp simulable from those that are not. A hint is provided by the nonlinear dependence on $\ann$ and $\adg$ seen in $\Lcal_2$ and $\Lcal_3$, suggesting that the physics separating paramp simulable and unsimulable amplifiers might have something to do with multiphoton processes. To tackle this question we focus on the family of counterexamples defined by $\Lcal_2$, which involve two-photon processes.

\begin{figure}[t]
\centerline{\includegraphics[width=0.47\textwidth]{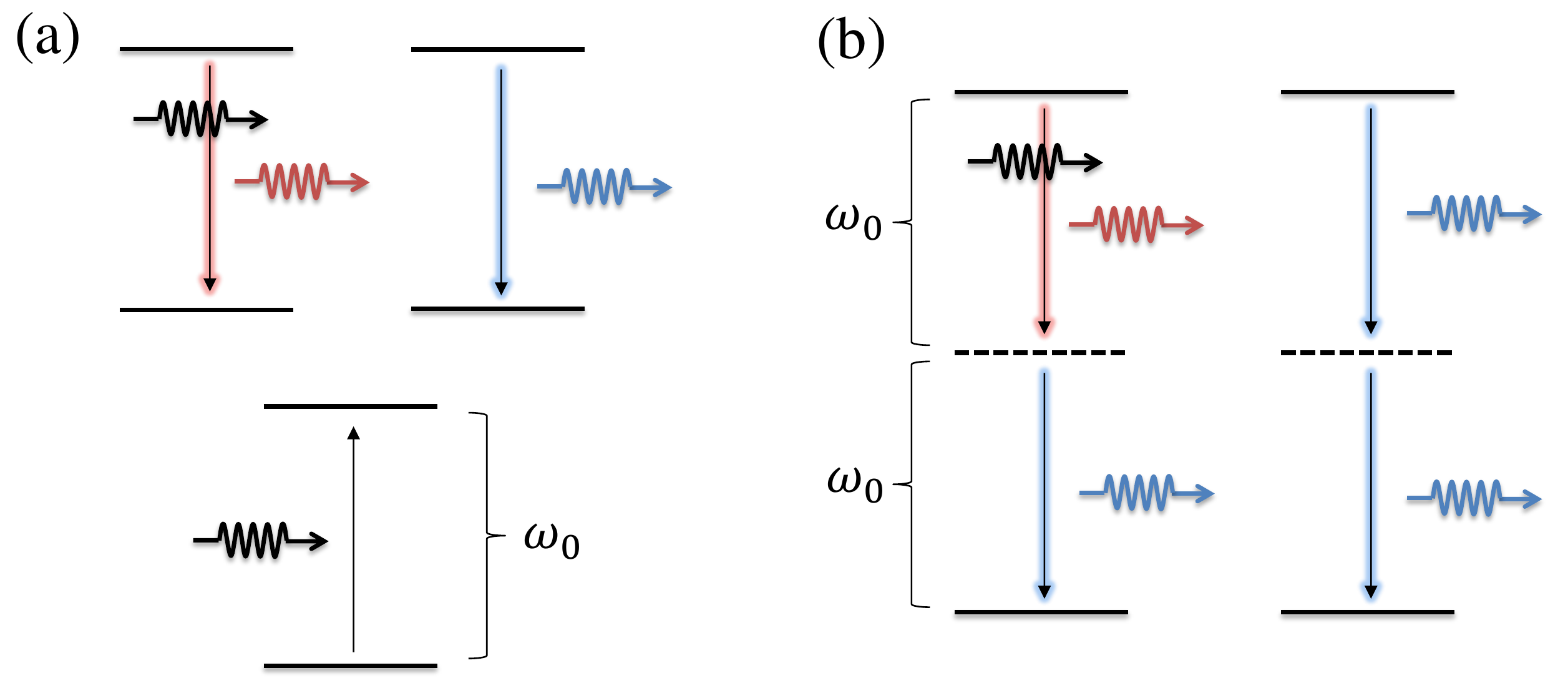}}
\caption{\label{AtomPhotonInt} Fundamental atom-photon interactions in $\Acal_1(t)$  and $\Acal_2(t)$. Photons emitted spontaneously (i.e.,~noise photons) are shown in blue while stimulated ones (i.e.,~signal photons) are in red (with the input signal photon shown in black). (a) $\Acal_1(t)$: Stimulated emission (top left), spontaneous emission (top right), and absorption (bottom).  (b) $\Acal_2(t)$: Left: Singly stimulated emission where a two-photon emission occurs as a cascade of stimulated and spontaneous emissions. Right: Two-photon spontaneous emission. Two-photon absorption and stimulated emission events (not shown) occur at the same rate when $\kupp = \kddn$ and do not provide net gain \cite{CHN+19}. }
\end{figure}

To start, we note that $\Lcal_2$ in fact appears as a special case of the so-called (degenerate) two-photon amplifier with the master equation \cite{Lam67,MW74}
\begin{align}
\label{TwoPhotonME}
	\ddt \, \rho(t) = \kupp \, \Dcal[\adg{}^2] \rho(t) + \kddn \, \Dcal[\ann^2] \rho(t)  \; .
\end{align}
This equation was derived from first principles starting from an atom-photon Hamiltonian with two-photon interactions by Lambropoulos in which $\kupp$ and $\kddn$ are further related to  microscopic quantities \footnote{Equation (3.13) of Ref.~\cite{Lam67} is equivalent to \eqref{TwoPhotonME} in the Fock basis.}. Here, it suffices to express them as $\kupp=\gamma\, n_\upp$ and $\kddn=\gamma\, n_\ddn$ where $\gamma$ is an effective atom-photon coupling strength while $n_\upp$ and $n_\ddn$ are the fractional atomic populations in the excited and ground states respectively. Two-photon amplifiers have been widely studied for some time \cite{Lam67,MW74,NEFE77,NZT81,BRGDH87,AGBZ90,GWMM92,Iro92,Gau03,NHO10,HNGO11,RSSHS16,MRBB18} and their output photon statistics have been intensively studied for the model of \eqref{TwoPhotonME} and special cases of it \cite{Lam67,MW74}. Already in Ref.~\cite{Lam67}, Lambropoulos noted that \emph{linear} amplification, i.e.~one-photon gain, was somehow possible with $\Lcal_2$ upon setting $\kupp=\kddn=\gamma/2$ in \eqref{TwoPhotonME} despite the amplifier being described by an inherently two-photon model [See Sec.~V.C of \cite{Lam67}. Also compare his Eqs.~(5.9b)--(5.9c) with our Eqs.~\eqref{<a>NoiseInduced}-\eqref{<n>NoiseInduced}]. To explain this he postulated that the amplification had to involve a ``half noise half signal'' process, originating from two-photon emissions whereby ``the emission of one of the photons is induced and the other spontaneous'' \footnote{See the main text on the last page of Ref.~\cite{Lam67}.}. However,  to the best of our knowledge, this assertion has remained unsubstantiated to date. If we are able to affirm the speculated mechanism underlying $\Lcal_2$, we would not only  have validated Lambropoulos's conjecture, but will also be guided to what kind of physics prevents a phase-preserving linear amplifier from being simulable by a paramp. As we now explain,  $\Lcal_2$ can be understood in terms of the elementary atom-photon interactions shown in Fig.~\ref{AtomPhotonInt}(b).

Attempts to understand the photon statistics of the two-photon amplifier naturally treat the density operator of the signal mode $\hat{a}$ as a central object of analysis, and thus work in the \sch\ picture. This is a major drawback in understanding the noise mechanism because the internal modes of the amplifier noise are traced out in such a description \cite{Lam67}. We are therefore motivated to work in the \hei\ picture where the amplifier noise appears explicitly as a time-dependent operator. This will allow us to track how the noise arises at the output  and arrive at Fig.~\ref{AtomPhotonInt}(b).

Before we analyse $\Acal_2(t)$ in the \hei\ picture, it is instructive to review how such an analysis works for the example of $\Acal_1(t)$. Its \hei-picture equivalent for the signal $\ann(t)$ can be shown to be given by \cite{CHN+19,GZ10}
\begin{align}
\label{da/dt=Sig+Noise}
	d\ann(t) = \frac{1}{2} \, \big( \kup - \kdn \big) \, \ann(t) \, dt + d\hat{W}(t)  \; ,
\end{align}
where $d\hat{s}(t) \equiv\hat{s}(t+dt)-\hat{s}(t)$ for  arbitrary $\hat{s}(t)$. Equation \eqref{da/dt=Sig+Noise} may be derived from a familiar model of the field interacting with a two-level atom. In this case,  $\kup$ and $\kdn$ are the effective excited-state and ground-state populations in an atomic gain medium that implements one-photon interactions. The term $d\hat{W}(t)$ is a quantum Wiener increment and represents the noise being added to the signal as it is being amplified according to \eqref{da/dt=Sig+Noise}. It is an atomic operator that is independent of the signal and has zero mean. All its higher-order moments vanish except the second-order ones given by the quantum \ito\ rules  \cite{GZ10,Ito42,Ito44,Ito46,HP84,WM10}
\begin{align}
	d\hat{W}\dg(t) \, d\hat{W}(t) = \kup \, dt \; ,  \quad   d\hat{W}(t) \, d\hat{W}\dg(t) = \kdn \, dt  \; .
\end{align}
Since we are now working explicitly in continuous time, the input and output signals are to be identified as $\ann(0)$ and $\ann(t)$ respectively. Applying quantum \ito\ calculus, \eqref{da/dt=Sig+Noise} can be shown to satisfy $[\ann(t),\adg(t)]=\hat{1}$ for all $t$, as required in order to be consistent with quantum mechanics.

The advantage of \eqref{da/dt=Sig+Noise} is that it  allows us to see how the noise contributes to the amplifier output explicitly. In particular, we can extract some basic physics about the amplification of $\ann$ by considering the evolution of the average photon number:
\begin{align}
\label{A1<dn>}
	d\ban{\hat{n}(t)} = {}& (\kup-\kdn) \, \ban{\hat{n}(t)} \, dt + d\hat{W}\dg(t) \, d\hat{W}(t)   \\
\label{StdLinAmpd<n>}
                                               = {}& (\kup-\kdn) \, \ban{\hat{n}(t)} \, dt + \kup \, dt  \; .
\end{align}
 The first two terms in \eqref{StdLinAmpd<n>} show that population inversion in the gain medium is necessary for a positive contribution to the signal's energy, i.e., for amplification. The third term given by $\kup$ comes directly from the noise operator  $d\hat{W}(t)$ and represents noise photons added to the signal. Furthermore, each term in \eqref{StdLinAmpd<n>} can be understood to correspond to an elementary atom-photon interaction (i.e.,~stimulated emission, absorption, or spontaneous emission) \cite{Aga13,Mil19}: The first term is proportional to both the intensity of the light reaching the atom $\an{\hat{n}(t)}$ as well as the effective atomic population of the excited state $\kup$ and corresponds to stimulated emission.  Similarly, we know that the number of absorption events in the gain medium should be proportional to $\an{\hat{n}(t)}$ and the effective ground-state population of the atoms. This corresponds to  the term $-\kdn\an{\hat{n}(t)}$ in \eqref{StdLinAmpd<n>} where the negative sign indicates that absorption removes energy from the field. The only atom-photon interaction that does not depend on the signal's energy, but only on the excited-state population of the gain medium, is spontaneous emission, and is given by the last term in \eqref{StdLinAmpd<n>}. This highlights the well-known facts about linear amplifiers that rely on single-photon interactions: First, that stimulated emission and population inversion are essential for amplification, and second, that spontaneous emission is the physical mechanism responsible for adding noise to the signal. A summary of these processes is shown in Fig.~\ref{AtomPhotonInt}(a).

The \hei-picture equation for $\ann$ corresponding to the two-photon amplifier of \eqref{TwoPhotonME} is \cite{CHN+19}
\begin{align}
\label{TwoPhotonIto}
	d\ann(t) = {}& \big( \kupp - \kddn \big) \, \ann\dg(t) \, \ann^2(t) \, dt    \nn \\
	                         & + 2\,\kupp \, \ann(t) \,dt + \ann\dg(t) \, d\hat{W}(t)  \, .
\end{align}
This is an \ito\ quantum stochastic differential equation \cite{Chi15,Par92,WZ65,Gou06,GC85} where $d\hat{W}(t)$ is again an atomic operator with zero mean and such that 
\begin{align}
	d\hat{W}\dg(t) \,d\hat{W}(t) = 4 \kupp\,dt \;,  \quad  d\hat{W}(t) \,d\hat{W}\dg(t) = 4 \kddn\,dt \; . 
\end{align}
Again, it can be shown that \eqref{TwoPhotonIto} preserves $[\ann(t),\adg(t)]=\hat{1}$ for all $t$  \cite{CHN+19}. The \hei\ equation of motion for $\ann$ corresponding to $\Lcal_2$ may then be obtained from \eqref{TwoPhotonIto} by setting $\kupp=\kddn=\gamma/2$. This gives
\begin{align}
\label{ItoFormNIA}
	d\ann(t) = \gamma \, \ann(t) \, dt + \adg(t) \, d\hat{W}(t)  \; .
\end{align}
Note here that \eqref{ItoFormNIA} now carries a signal-dependent noise given by $\adg(t) d\hat{W}(t)\,$. This is the ``half signal half noise'' which Lambropoulos spoke of in Ref.~\cite{Lam67}. It is also referred to as multiplicative noise in random process theory \cite{Jac10,Gar09}. We can now show exactly what the multiplicative noise in \eqref{ItoFormNIA} is in terms of elementary atom-photon interactions by considering how the average photon number evolves. Using quantum \ito\ calculus we have,
\begin{align} 
	d\ban{\hat{n}(t)} = {}& 2 \,\gamma  \ban{\nhat(t)} \, dt  + \ban{ \ann(t) \,\adg(t) } \, d\hat{W}\dg\!(t) \, d\hat{W}(t)   \\
\label{dnItoProduct}
                                               = {}& 2 \,\gamma  \ban{\nhat(t)} \, dt  + 2\,\gamma \, \big[ \ban{ \nhat(t)} + 1 \big] \, dt  \; .
\end{align}
The first term in \eqref{dnItoProduct} is inherited from the $\gamma\,\ann(t)dt$ term in \eqref{ItoFormNIA} and corresponds to one-photon stimulated emission as it depends on $\kupp$ and $\an{\nhat(t)}$. Since the model restricts the  atoms to have only two-photon transitions, this term by itself does not complete a full atomic transition from excited to ground state with the emission of two photons.  To complete the picture we must take into account for the photons from the remaining terms in \eqref{dnItoProduct}, which are noise photons insofar as they arise from the atomic operator $d\hat{W}(t)$. In contrast to \eqref{StdLinAmpd<n>}, there are now two types of noise photons. The first is linear in $\an{\nhat(t)}$, so it  corresponds to a one-photon emission that depends on the signal strength reaching the atom. The fact that it is a noise photon suggests that it came from spontaneous emission while the fact that it depends on the signal means that such a spontaneous emission is ``stimulated''---conditioned on a stimulated emission having taken place just before it. The seemingly strange possibility of getting one-photon amplification in a two-photon model can now be resolved when we take the stimulated photon corresponding to the first term in \eqref{dnItoProduct} together with  the signal-dependent noise photon to arrive at the two-photon process shown on the left of Fig.~\ref{AtomPhotonInt}(b). This is the underlying mechanism responsible for linear (i.e.~one-photon) amplification in a gain medium with only two-photon transitions.  The remaining type of noise photon is due to the $2\gamma$ in \eqref{dnItoProduct} which corresponds to two-photon spontaneous emission. This is shown on the right in Fig.~\ref{AtomPhotonInt}(b).

Our physical picture of the multiplicative noise in \eqref{ItoFormNIA} thus allows us to see how it is signal dependent. It is precisely this signal-dependent noise that leads to a photon-number gain of $g^4$ in \eqref{<n>NoiseInduced} which ultimately makes it impossible for the paramp to simulate it as shown in \eqref{<n>1-<n>2}. This can  be seen explicitly from \eqref{dnItoProduct} where the first term contributes  photons at a rate $2\gamma\an{\nhat(t)}$  to the signal, while the signal-dependent noise contributes another $2\gamma\an{\nhat(t)}$ photons per unit time to make up a total rate of $4\gamma\an{\nhat(t)}$ [which leads to the fourth power of $g$ in \eqref{<n>NoiseInduced} and subsequently in \eqref{<n>1-<n>2}]. Because \eqref{ItoFormNIA} is the simplest form of a phase-preserving linear amplifier with multiplicative noise, it may be expected  that other such amplifiers with more complicated signal-dependent noise  can also violate \eqref{ParampConj}, as we showed with $\Acal_3(t)$ from Eq.\eqref{ThreePhotonGenerator}.

We note that non-degenerate variants of the left picture in Fig.~\ref{AtomPhotonInt}(b) (i.e.~a two-photon emission with unequal transition frequencies) has been observed in experiments and are known in the literature as singly stimulated emission \cite{HGO08,NBH+10,OIKA11} (see Ref.~\cite{HNGO11} and the references therein for more details). What we have done in this section on the physical properties of our counterexamples is to show that (i) multiplicative noise prevents a phase-preserving linear amplifier from being paramp simulable, and (ii) explain the physical basis of this multiplicative noise in terms of elementary atom-photon interactions.

It is also possible to interpret \eqref{ItoFormNIA} and its associated linear amplification purely from the perspective of quantum stochastic processes. In this interpretation \eqref{ItoFormNIA} is understood to generate linear amplification as a result of the correlations between the amplifier-added noise and the signal. This follows from the \str\ form of \eqref{ItoFormNIA} which is derived in Ref.~\cite{CHN+19}. Such a process may in principle be realised using ion traps \cite{SMrefs}.

Finally, our discussion above sheds light on how $\Acal_2$ evades the claimed proof of the universality of the paramp model in Ref.~\cite{CCJP12}. The authors of Ref.~\cite{CCJP12} mathematically characterize a phase-preserving linear amplifier as a composition of a perfectly noiseless (and unphysical) amplifier with a noise map that restores physicality \footnote{See  Equation (3.2) and the surrounding discussion in Ref.~\cite{CCJP12}.}. Crucially, this added noise was taken to be \emph{signal-independent}, thus excluding multiplicative noise of the kind found in $\Acal_2$ by fiat.

\begin{acknowledgments}

We would like thank Carl Caves for email correspondences and Howard Wiseman for some feedback on our paper draft. In addition we thank Berge Englert, Christian Miniatura, Alex Hayat, Aaron Danner, and  Tristan Farrow for useful discussions on atom-photon interactions. This research is supported by: The MOE grant number RG 127/14, the National Research Foundation, Prime Minister's Office, Singapore under its Competitive Research Programme (CRP Award No. NRF-CRP-14-2014-02), the National Research Foundation of Singapore (NRF Fellowship Reference Nos. NRF-NRFF2016-02 and NRF-CRP14-2014-02), the Ministry of Education Singapore (MOE2019-T1-002-015), the National Research Foundation Singapore and the Agence Nationale de la Recherche (NRF2017-NRFANR004 VanQuTe), and the Foundational Questions Institute (FQXi-RFP-IPW-1903). MH acknowledges support by the Air Force Office of Scientific Research under award FA2386-19-1-4038.

\end{acknowledgments}


\onecolumngrid
\newpage

\section*{\large Supplementary Material for ``Phase-Preserving Linear Amplifiers Not Simulable by the Parametric Amplifier''}

\section{The two-photon amplifier}

\subsection{Overview}

We have already said in the main text that the degenerate two-photon amplifier can be modelled by the master equation
\begin{align}
\label{TwoPhotonMESuppMat}
	\ddt \; \rho(t) = \kddn \, \Dcal[\ann^2] \, \rho(t) + \kupp \, \Dcal[\adg{}^2] \, \rho(t)  \; .
\end{align}
This may be derived within the Born--Markov framework of open-systems theory \cite{DH14,BP02SM,Car02SM}. The procedure leading to the master equation \eqref{TwoPhotonMESuppMat} is fairly well understood so we will not derive it here. We will, however, derive the corresponding \hei\ equation of motion for $\ann$ in Sec.~\ref{HeiPic} since it is the \hei-picture treatment that has played a critical role for understanding the physics of the multiplicative noise which we met in \eqref{TwoPhotonIto}--\eqref{dnItoProduct}. In addition, the \hei-picture treatment of open systems is somewhat less well known compared to the \sch-picture theory. The specific form of the \hei\ equation of motion used in \eqref{TwoPhotonIto}--\eqref{dnItoProduct} assume that the atomic baths behave as white noise so we must then take this limit after deriving the \hei\ equation of motion. This then turns the \hei\ equation for $\ann$ into a quantum stochastic differential equation.

Generally, a stochastic differential equation can be classified to be one of two kinds \cite{Gar09, Jac10a}: The first kind is a stochastic differential of the \str\ form. The second kind is a stochastic differential equation of the \ito\ form. One advantage of the \str\ form is that normal calculus can used when manipulating these equations. However, this makes the statistical moments of $\ann$, such as the photon number, or two-time correlation functions more cumbersome to derive. On the other hand, a stochastic differential equation in the \ito\ form requires one to learn new rules of differentiation and integration. This is known as \ito\ calculus. \ito\ equations have the advantage that its noise terms are always independent of the system variables and this leads to computational simplicity provided that \ito\ calculus is correctly applied. These general properties of \str\ and \ito\ calculi also apply to quantum stochastic differential equations \cite{Par92,GZ10SM}.

The difference between the two forms of stochastic differential equations originate in the order in which the white-noise limit is taken. We obtain a \str\ quantum stochastic differential equation when we take the white-noise limit of the \hei\ equation of motion at the end of its derivation (as opposed to the start). The rigorous justification of this is given by the Wong--Zakai theorem \cite{WZ65SM,Gou06SM}. Hence, when we take the white-noise limit of the resulting \hei\ equation of motion for $\ann$ in Sec.~\ref{HeiPic} we arrive at a \str\ equation. From this we will proceed to derive the average evolution from this within the \str\ framework in Sec.~\ref{NoisiAmplification}. This calculation allows us to understand linear amplification in \eqref{TwoPhotonMESuppMat} as the result of correlations between the noise and the signal. For this reason one may refer to the case of $\kupp=\kddn$ as a noise-induced amplifier. As just mentioned, the \str\ form of the \hei\ equation makes it more difficult to calculate moments of $\ann$, so we will convert our \str\ quantum stochastic differential equation for $\ann$ into its equivalent \ito\ form. This then gives us exactly \eqref{TwoPhotonIto} in the main text. We then show, in Sec.~\ref{ProofCCR}, that on using \ito\ calculus the canonical commutation relation $[\ann(t),\adg(t)]=\hat{1}$ is preserved for all $t$ as it should be in the \hei\ picture.

\subsection{Amplitude equation of motion}
\label{HeiPic}

The two-photon amplifier given in the main text by the master equation \eqref{TwoPhotonME} and the \ito\ stochastic equation \eqref{TwoPhotonIto} may be derived by modelling the signal as a single bosonic oscillator (with Hilbert space $\mathbbm{H}_A$) coupled to a bath of two-level atoms (with Hilbert space $\mathbbm{H}_B$) that mediate two-photon transitions. The atoms model the gain medium that is used for amplification. The full Hamiltonian $\Hhat$ on $\mathbbm{H}_A \otimes \mathbbm{H}_B$ is
\begin{align}
\label{HamiltonianHP}
	\Hhat = \hbar \, \wo \, \adg \ann + \sum_n \frac{\hbar\,\omega_n}{2} \; \piez_n + \annt \, \Pihat\dg + \adgt \, \Pihat \; , 
\end{align}
where $\omega_0$ is the oscillator's natural frequency and $\piez_n$, $\pieplus_n$, and $\piemin_n$ are atomic operators for the $n$th atom, defined by
\begin{align}   
	 \piez_n = \pieplus_n \piemin_n - \piemin_n \pieplus_n  \;, \quad   \pieplus_n = \op{\upp_n}{\ddn_n}  \;, \quad   \piemin_n = \op{\ddn_n}{\upp_n}  \;.
\end{align}
We have also defined the bath operators
\begin{align}
	\Pihat = \hbar \sum_n \, \xi_n \, \piemin_n  \;,  \quad \Pihat\dg = \hbar \sum_n \xi^*_n \, \pieplus_n  \;.
\end{align}
The bath will be assumed to be at temperature $T$ so that its state $\rho_B$ is given by
\begin{align}
	\rho_B = \bigotimes_n \, \rho_{n}  \;,   \quad   \rho_{n} = \frac{\exp\!\big(\!-\!\beta\hbar\;\!\omega_n \piez_n/\;\!2\big)}{Z_{n}}  \;,
\end{align}
where we have defined $\beta= 1/\kB\,T$, and $\kB$ is the Boltzmann constant. The normalisation of $\rho_n$ (also the partition function) is
\begin{align}
	Z_{n} = {\rm Tr}\big[\exp\!\big(\!-\!\beta\hbar\;\!\omega_n \piez_n/\;\!2\big)\big] = 2 \cosh\!\bigg( \frac{\beta \hbar\;\!\omega_n}{2} \bigg)  \;.
\end{align}
It will also be useful to introduce the shortands for the atomic populations in the $n$th atom:
\begin{align}
	\bra{\upp_n} \rho_n \ket{\upp_n} \equiv N_{\upp}(\omega_n,T)  \;,  \quad   \bra{\ddn_n} \rho_n \ket{\ddn_n} \equiv N_{\ddn}(\omega_n,T)  \;.
\end{align}
%

%
%
%

The \hei\ equation of motion for the oscillator's amplitude $\ann$ is defined by
\begin{align}
\label{da/dt}
	\frac{d}{dt} \, \ann(t) = -\frac{i}{\hbar} \; e^{i\hat{H}t/\hbar} \big[ \ann(0), \Hhat(0) \big] e^{-i\hat{H}t/\hbar}  \;,
\end{align}
where $\Hhat(0)=\hat{H}$. Noting that at the initial time the system and bath operators commute, we find
\begin{align}
\label{da/dt2}
	\frac{d}{dt} \, \ann(t) = -i \, \wo \, \ann(t) - \frac{i}{\hbar} \, 2 \, \adg(t) \, \Pihat(t)    \;.
\end{align}
where $\Pihat(t)=\exp(i\hat{H}t/\hbar)\,\Pihat\,\exp(-i\hat{H}t/\hbar)$. It helps to move into a rotating frame at the oscillator frequency by defining
\begin{align}
\label{aRF}
	\abar(t) = \ann(t) \, e^{i\,\wo t}  \;.
\end{align}
Differentiating $\abar(t)$ and using \eqref{da/dt2} we get
\begin{align}
\label{da/dt3}
	\frac{d}{dt} \, \abar(t) = - \frac{i}{\hbar} \, 2 \, \abar\dg(t) \, \Pihat(t) \, e^{i\,2\wo\,t}  \;.
\end{align}
We see that $\abar(t)$ is coupled to $\Pihat(t)$. To deal with this one may substitute the formal solution for $\Pihat(t)$ back into \eqref{da/dt3} iteratively. However, if the system and bath are only weakly coupled then we can approximate the system evolution up to second order in the interaction strength. This step constitutes the so-called Born approximation, after which we arrive at   
\begin{align}
\label{dabar/dtBorn}
	\frac{d}{dt} \, \abar(t) = - \frac{i}{\hbar} \, 2 \, \abar\dg(t) \, \Pip(t) \, e^{i \,2\wo t} 
	                           - \frac{2}{\hbar^2} \, \abar\dg(t) \int^t_0 dt' \: \abar^2(t') \, \big[ \Pip(t), \Pip\dg(t') \big]  \, e^{i\,2\wo(t-t')}  \;,
\end{align}
where we have defined
\begin{align}
\label{IntPicPi}
	\Pip(t) = \hbar \sum_n \, \xi_n \, \piemin_n(0) \, e^{-i\,\omega_n t}  \;.
\end{align}
The system's evolution is now affected by the history of $\Pip(t)$ in the commutator inside the integrand. We can simplify this by first replacing the bath commutator by its average, which is justified if we are going to use the \hei\ equation of motion for calculating expectation values. The dependence on the history of $\Pip(t)$ can then simplified by making the Markov approximation: This relies on the characteristic timescale over which $\abar(t)$ evolves to be much longer than the timescale over which bath correlations decay. In this regime we can then replace the system operators at time $t'<t$, by the present time $t$ and extend the top limit of the time integrals to infinity. Doing so allows us to compute the time integral in \eqref{dabar/dtBorn} by assuming the distribution of transition frequencies of the atoms to be sufficiently dense. We may then convert the sum over atomic degrees of freedom in $\Pip(t)$ into an integral by introducing a function $D(\omega)$ which counts how many atoms there are per transition frequency in the bath. That is, $D(\omega)\,d\omega$ is the number of atoms in the bath with a transition frequency in the range from $\omega$ to $\omega+d\omega$. We then have
\begin{align}
\label{RFda/dt}
	\frac{d}{dt} \, \abar(t) = \big( \kupp - \kddn \big) \, \abar\dg(t) \, \abar^2(t) - \frac{i}{\hbar} \, 2 \, \abar\dg(t) \, \Pip(t) \, e^{i \,2\wo t}    \;,
\end{align}
where we have further defined
\begin{align}
	\kupp = \gamma \, N_{\upp}(2\wo,T) \;,  \quad  \kddn = \gamma \, N_\ddn(2\wo,T) \;, \quad  \gamma \equiv 2 \;\!\pi \, D(2\wo) \, |\xi(2\wo)|^2  \;.
\end{align}
In \eqref{RFda/dt} we have neglected shifts in the oscillator's frequency due to the bath correlation functions on the grounds that they are typically very small \cite{BP02SM,Car02SM}. Physically this can be expected since atoms that are detuned from $2\wo$ would not be expected to have a strong two-photon coupling. The dominant coupling occurs for the on-resonance case and they give rise to $\kupp$ and $\kddn$.

\subsection{Noise-induced amplification}
\label{NoisiAmplification}

\subsubsection{Effect of noise within \str\ calculus}

Taking the expectation value of \eqref{RFda/dt} gives us 
\begin{align}
\label{<RFda/dt>}
	\frac{d}{dt} \, \ban{\abar(t)} = \big( \kupp - \kddn \big) \, \ban{\abar\dg(t) \, \abar^2(t)} - \frac{i}{\hbar} \, 2 \, \ban{\abar\dg(t) \, \Pip(t)} \, e^{i \,2\wo t}    \;,
\end{align}
If we can calculate the expectation value of the noise term in \eqref{<RFda/dt>} then we know what effect it has on the signal on average. However, it is not immediately obvious how to do this since the evolution of $\abar(t)$ under \eqref{RFda/dt} will correlate it with $\Pip(t)$. Therefore we must treat $\ban{\abar\dg(t) \, \Pip(t)}$ with care. Typically it is difficult to proceed further without assuming anything about $\Pip(t)$. Therefore it is often useful to consider the white-noise limit of \eqref{RFda/dt}. In so doing we obtain the \str\ equivalent to \eqref{TwoPhotonIto} from the main text (except for some scaling of the noise operator which we will take into account later). This means that we may approximate the autocorrelations of $\tilde{\Pi}(t)$ to have as small a correlation time as we like. Effectively one may take
\begin{align}
	\ban{\Pip\dg(t) \, \Pip(t-\tau)} = \frac{\hbar^2}{2} \; \kupp \, \delta(\tau) \; ,  \quad    \ban{\Pip(t) \, \Pip\dg(t-\tau)} = \frac{\hbar^2}{2} \; \kddn \, \delta(\tau)  \; .
\end{align}
Integrating \eqref{RFda/dt} allows us to arrive at
\begin{align}
\label{<sys(t)noise(t)>1}
	\ban{\abar\dg(t)\;\!\Pip(t)} = \ban{\abar\dg(0)\;\!\Pip(t)} + \big( \kupp - \kddn \big) \int^t_0 dt' \, \ban{\abar\dg{}^2(t')\,\abar(t')\;\!\Pip(t)} 
	                               + 2\,\frac{i}{\hbar} \int^t_0 dt' \, \ban{\Pip\dg(t')\;\!\abar(t')\;\!\Pip(t)} \, e^{-i \,2\wo t'}  \;.
\end{align}
Because we are assuming $\Pip(t)$ to have short correlation times we can factorise multitime averages between noise operators at time $t$ and system operators at time $t'$ provided that $t>t'$. For example, for any system operator $\hat{s}$, 
\begin{align}
\label{SysNoise1}
	\ban{\hat{s}(t')\,\Pip(t)} = \ban{\hat{s}(t')} \ban{\Pip(t)} = 0  \;,   \quad \; t>t'  \; ,
\end{align}
where we have noted that $\Pip(t)$ has zero mean. When $t>t'$ the noise $\Pip(t)$ and system variable $\hat{s}(t')$ are in fact independent so we have
\begin{align}
\label{SysNoise2}
	\big[ \hat{s}(t'),\Pip(t) \big] = 0  \;,   \quad \; t>t'  \; .
\end{align}
The case of $t=t'$ then depends on the form of $\hat{s}$ and the operator that couples to the bath as defined by $\Vhat$  \cite{GC85}. For the case considered here this can be seen to give $[\abar(t),\Pip(t)]=0$. Hence we have,
\begin{align}
\label{<sys(t)noise(t)>2}
	\ban{\abar\dg(t)\;\!\Pip(t)} 
	= 2 \, \frac{i}{\hbar} \, \an{\abar(t)} \, e^{-i\,2\wo \;\!t}  \int^\infty_0 d\tau \, \ban{\Pip\dg(t-\tau)\;\!\Pip(t)} \: e^{i\,2\wo \tau}   
	= i \;\! \hbar \, \kupp \, \an{\abar(t)} \, e^{-i\,2\wo \;\!t}  \;.
\end{align}
Substituting this back into \eqref{<RFda/dt>} thus gives
\begin{align}
\label{RF<aPi>}
	\frac{d}{dt} \, \an{\abar(t)} = \big( \kupp - \kddn \big) \, \ban{\abar\dg(t) \, \abar^2(t)} +  2\,\kupp \, \an{\abar(t)}  \;.  
\end{align}
For ease of writing let us define
\begin{align}
\label{w(Pi)}
	\hat{w}(t) = - \frac{i}{\hbar} \, 2 \, \Pip(t) \, e^{i \,2\wo t}  \; ,
\end{align}
and relabel $\abar(t)$ as $\ann(t)$ (keeping in mind that it is an equation of motion in the rotating frame). Considering the case of $\kupp=\kddn=\gamma/2$ we can then write \eqref{RFda/dt} simply as
\begin{align}
\label{NoisiampStr}
	\frac{d}{dt} \, \ann(t) = \adg(t) \, \hat{w}(t) \; ,
\end{align}
and where the average of this is simply \eqref{RF<aPi>} written in terms of $\hat{w}(t)$,
\begin{align}
\label{NoisiampAvg}
	\ban{\adg(t) \, \hat{w}(t)} = \gamma \, \an{\ann(t)}  \; .
\end{align}

The derivation of \eqref{NoisiampAvg} proves that linear amplification can be induced by the amplifier added noise when it comes in the form of multiplicative noise. For this reason we can refer to \eqref{NoisiampStr} as a noise-induced amplifier (henceforth abrreviated to noisiamp). We can also understand noisi amplification as classical correlation between the internal noise source of the amplifier and the signal that is being amplified. This is the essential content of \eqref{NoisiampAvg}. Though such equations are not typically encountered in the amplifier literature, it is certainly allowed within the Born--Markov framework of open-systems theory. The important point to note here is that neither the Markov approximation, nor \str\ calculus, treat $\hat{w}(t)$ as true idealised white noise. All that is required is for $\hat{w}(t)$ to have a very small but nonzero correlation time, otherwise \eqref{NoisiampAvg} would be zero and there would be no point to the derivation above. In other words, if there is no correlation between the noise and signal, there is no amplification. Having said this, it is possible to convert \eqref{NoisiampStr} to a form where $\hat{w}(t)$ really is ideal white-noise for which its correlations with any system variable always vanishes. This is given by the \ito\ form corresponding to \eqref{NoisiampStr} and we will consider it in Sec.~\ref{ProofCCR}. Before taking this on we briefly discuss how one might be able to observe noisi amplification in ion traps.

\subsubsection{Realisation using ion traps}

The noisiamp in \eqref{NoisiampStr} and \eqref{NoisiampAvg} may in principle be realised using ion traps as follows. Trapped ions can be thought of as possessing two degrees of freedom, an internal degree of freedom which we can effectively think of as a two-level atom, and a motional degree of freedom. The internal degree of freedom has basis states $\ket{g}$ (ground state) and $\ket{e}$ (excited state), while the Fock basis $\ket{n}$ is used for the motional degree of freedom. Implementing \eqref{NoisiampStr} and \eqref{NoisiampAvg} is equivalent to implimenting \eqref{TwoPhotonMESuppMat} with $\kupp=\kddn=\gamma/2$. 

\begin{figure}[b]
\centering
\includegraphics[scale=0.8]{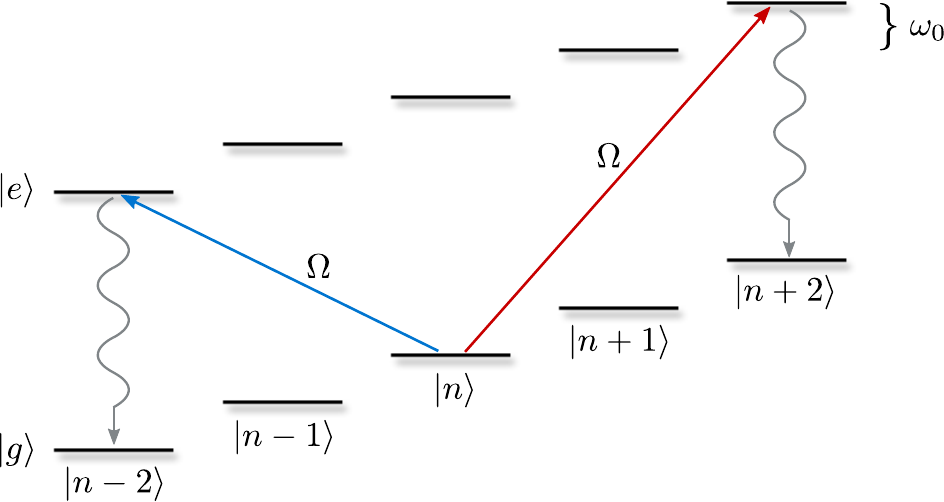}
\caption{\label{IonTrap} Energy diagram of a trapped ion with natural frequency $\omega_0$ interacting with two laser fields detuned by $\pm\omega_0$ from the carrier frequency. Relaxation processes occur at a much faster rate compared to the sideband interactions, effectively implementing two-photon heating and cooling.}
\end{figure}
In order to implement the two-photon cooling in \eqref{TwoPhotonMESuppMat} with $\kupp=\kddn=\gamma/2$, the trapped ion interacts with a laser field detuned from the carrier transition by $-2\omega_0$, where $\omega_0$ is the natural frequency of the trap. The ion then relaxes at the carrier frequency, effectively implementing two-photon loss ($\ket{g,n} \rightarrow \ket{g,n-2}$). Similarly, another laser field detuned by $2\omega_0$ is used to implement the two-photon heating process ($\ket{g,n} \rightarrow \ket{g,n+2}$). These processes are illustrated in Fig.~\ref{IonTrap}. The ion must be deep in the Lamb-Dicke regime \cite{leibfried2003quantum}, ensuring the sidebands are resolved and relaxation occurs predominantly at the carrier frequency. The Lamb--Dicke parameter is given by $\eta=(2\pi/\lambda)\sqrt{\hbar/2m\omega_0}\,\cos\theta$, where $\lambda$ is the wavelength of the incident laser field, $m$ is the mass of the ion, and $\theta$ is the angle between the laser field and the motion of the ion. The Lamb-Dicke parameter must satisfy $\eta^2(2n+1)\ll 1$, where $n$ is the Fock state of the ion's motion. To implement \eqref{TwoPhotonMESuppMat} with $\kupp=\kddn=\gamma/2$, the heating and cooling is required to have the same rate $\gamma/2$, which is controlled by the Rabi frequency $\Omega$ of the applied laser fields, and is given by $\gamma\approx\eta^2\Omega$. A similar realisation has been proposed in Ref.~\cite{lee2013quantum} to implement the quantum van der Pol oscillator.

\subsection{Conversion to \ito\ form and consistency with quantum mechanics}
\label{ProofCCR}

\subsubsection{The two-photon amplifier in \ito\ form}

Using \eqref{w(Pi)} we may write the \str\ equation \eqref{RFda/dt} as
\begin{align}
\label{da2PhStr}
	\frac{d}{dt} \; \ann(t) = \big( \kupp - \kddn \big) \, \adg(t) \, \ann^2(t) + \adg(t) \, \hat{w}(t)  \; .
\end{align}
From \eqref{<sys(t)noise(t)>2} it is not difficult to see that the \ito\ equivalent of \eqref{da2PhStr} is given by 
\begin{align}
\label{da2PhIto}
	d\ann(t) = \big( \kupp - \kddn \big) \, \ann\dg(t) \, \ann^2(t) \, dt  + 2\,\kupp \, \ann(t) \,dt  + \ann\dg(t) \, d\hat{W}(t)  \; ,
\end{align}
where $d\hat{s}(t)\equiv\hat{s}(t+dt)-\hat{s}(t)$ for any operator $\hat{s}$ and $d\hat{W}(t)=\hat{w}(t)\,dt$ satisfies the \ito\ rules
\begin{align}
\label{ItoRulesAppendix}
	d\hat{W}\dg(t) \, d\hat{W}(t) = 4 \, \kupp \, dt \;,  \quad  d\hat{W}(t) \, d\hat{W}\dg(t) = 4 \, \kddn \, dt  \; .
\end{align}
Note that as we have mentioned earlier, the \ito\ equation has the property that
\begin{align}
	\ban{\ann\dg(t)\,d\hat{W}(t)} = 0 \; . 
\end{align}
However, there is now an extra term of order $dt$ in \eqref{da2PhIto} containing $2\,\kupp \, \ann(t)$ that makes sure it has the same average as \eqref{da2PhStr}. While the \str\ and \ito\ equations can have different appearances, the physics described by each within their respective calculus must be identical. Equation \eqref{da2PhIto} can be derived directly from \eqref{da2PhStr} by treating the time derivative as an implicit equation \cite{WM10SM}. Alternatively, \eqref{da2PhIto} can also derived by treating $\hat{w}(t)$ as a quantum white-noise process from the start. This can be achieved using the time-evolution operator in the rotating frame:
\begin{align}
\label{ItoU(t,0)}
	\hat{U}(t,0) = {\rm T}_\triangleleft \bigg\{ \exp\bigg[ \!-\!i \int^t_0 dt' \, \Vhat(t') \bigg] \bigg\}  \; ,
\end{align}
where ${\rm T}_\triangleleft$ denotes chronological time ordering. Here $\Vhat(t)$ is in the rotating frame with respect to the free evolution and where all such time dependencies are grouped into $\hat{w}(t)$ [see \eqref{w(Pi)}],
\begin{align}
\label{V(t)ItoDefn}
	\Vhat(t) = \ann^2(0) \, \hat{w}\dg(t) + \adg{}^2(0) \, \hat{w}(t). 
\end{align}
The time dependence of $\ann(t)$ in \eqref{da2PhIto} is thus defined by 
\begin{align}
\label{a(t)ItoDefn}
	\ann(t) = \hat{U}\dg(t,0) \, \ann(0) \, \hat{U}(t,0) \; .
\end{align}
Often the time-evolution operator is specified by a Hudson--Parthasarthy equation (a quantum stochastic \sch\ equation) \cite{HP84,Chi15,WM10SM}
\begin{align}
\label{HPE}
	d\hat{U}(t,0) = \bigg\{ \!-\!\frac{1}{2} \; \Big[ \, \kupp \, \ann^2(t) \, \adg{}^2(t) + \kddn \, \adg{}^2(t) \, \ann^2(t) \Big] \, dt  
	                              - \frac{1}{2} \, \Big[ \ann^2(t) \, d\hat{W}\dg(t) - \adg{}^2(t) \, d\hat{W}(t)  \Big] \bigg\} \, \hat{U}(t,0)  \; .
\end{align}
In practice it is easier to derive the \ito\ quantum stochastic differential equation from the Hudson--Parthasarathy/stochastic \sch\ equation \eqref{HPE}. We emphasise here that \eqref{ItoU(t,0)} [or \eqref{HPE}], along with \eqref{V(t)ItoDefn} and \eqref{ItoRulesAppendix} provide an independent and way of deriving \eqref{da2PhIto} that is void of any reference to baths at thermal equilibrium. Equation \eqref{V(t)ItoDefn} simply couples the signal represented by $\ann(t)$ to a quantum white-noise process $\hat{w}(t)$.

\subsubsection{Preservation of canonical commutation relation}

\begin{figure}[t]
\centerline{\includegraphics[width=0.5\textwidth]{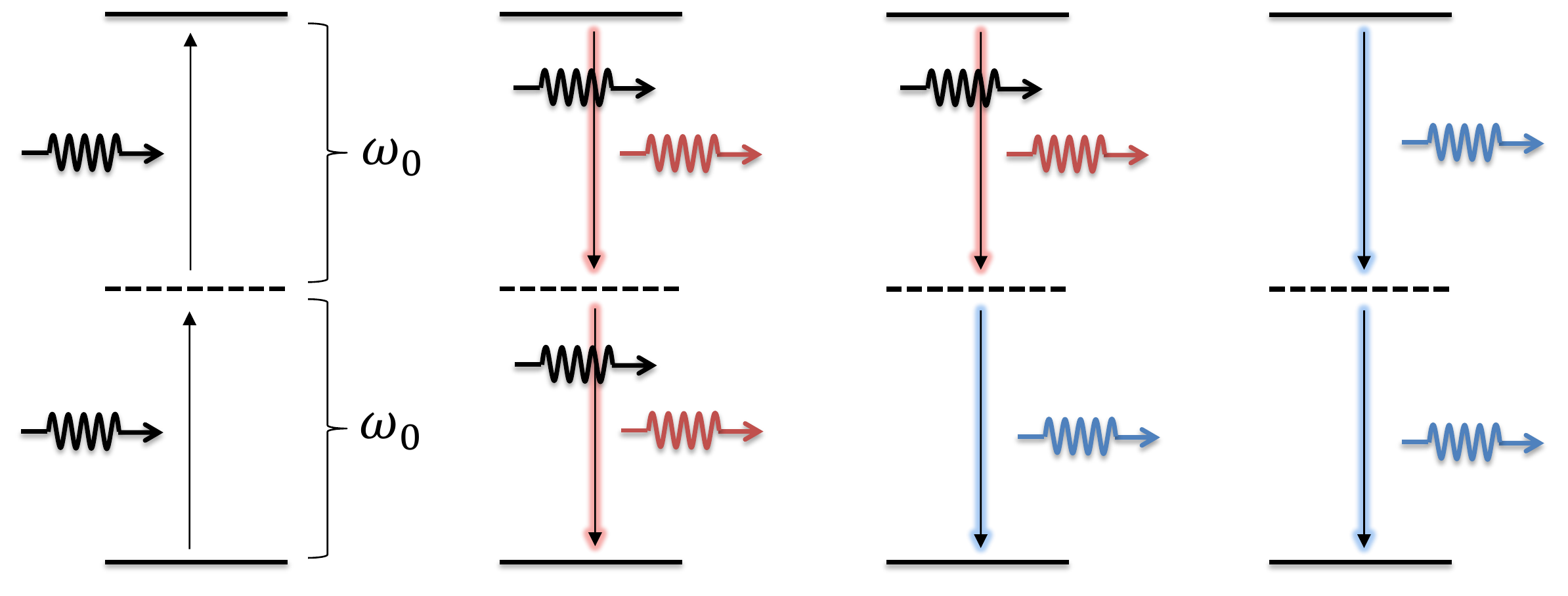}}
\caption{\label{TwoPhotonProcessSuppMat} Various two-photon atom-field interactions occurring in \eqref{d<adga>/dt} via an intermediate virtual level (dashed line). From left to right: Two-photon absorption, two-photon stimulated emission, two-photon emission with one stimulated emission and one spontaneous, two-photon spontaneous emission.}
\end{figure}

We can show that \eqref{da2PhIto} preserves the canonical commutation relation for $\ann$ and $\adg$. If at time $t$ we have $[\ann(t),\adg(t)]=\hat{1}$, then we must have
\begin{align}
\label{d[a,adg]}
	d\big[\ann(t),\adg(t)\big] = 0   \; .
\end{align}
Omitting the time argument for ease of writing we have
\begin{align}
	d\big[\ann,\ann\dg\big] 
	= {}& (d\ann) \, \ann\dg + \ann \, (d\ann\dg) + (d\ann) (d\ann\dg) - (d\ann\dg) \ann - \ann\dg (d\ann) - (d\ann\dg) (d\ann)   \\
	= {}& (\kupp-\kddn) (\ann\dg \ann^2 \ann\dg + \ann \, \ann\dg{}^2 \ann - 2 \, \ann\dg{}^2 \ann^2 ) \, dt - 4 \, ( \kupp-\kddn) \, \ann\dg \ann \, dt  \;.
\end{align}
On normal ordering the first two terms in the parentheses on the right-hand side we arrive at \eqref{d[a,adg]}. As part of the proof of \eqref{d[a,adg]} we have also worked out the photon-number evolution in the general case when $\kupp\ne\kddn$. Its average gives
\begin{align}
\label{d<adga>/dt}
	\frac{d}{dt} \; \an{\adg\ann} = 2 \, (\kupp-\kddn) \, \ban{\adgt\annt} + 8 \, \kupp \, \ban{\adg\ann} + 4 \, \kupp  \;.
\end{align}
In the main text we worked out the corresponding atom-photon interactions taking place when $\kupp=\kddn$ so that the nonlinear term in \eqref{d<adga>/dt} does not contribute to the noisiamp [see Fig.~\ref{AtomPhotonInt}(b) of the main text]. The nonlinear term here represents a two-photon generalisation of the linear (i.e.~one-photon) amplifier. We depict the necessary atom-photon interactions associated with the general two-photon amplifier in Fig.~\ref{TwoPhotonProcessSuppMat}.

\section{Phase properties}
\label{PhaseCovarianceSM}

\subsection{Phase covariance}

We define an arbitrary linear amplifier to be phase covariant if and only if its map $\Acal(t)$ (assumed to be completely-positive and trace-preserving) commutes with the phase-shift map,
\begin{align}
\label{PhaseCovDefn}
	\Acal(t) \, \Phase = \Phase \, \Acal(t)  \; .
\end{align}
The phase-shift map is defined by 
\begin{align}
	\Phase \, \rho = \exp(-i\varphi\hat{n}) \, \rho \, \exp(i\varphi\hat{n}) \equiv \rho_\varphi  \; .
\end{align}
where $\nhat=\adg\ann$. This same property has been referred to as ``phase-preserving in the strict sense'' in Ref.~\cite{CCJP12SM}.

To prove that $\Acal(t)$ is phase covariant, we can think of $\Acal(t)$ as many compositions of $\Acal(\delta t)$ in the limit that $\delta t \longrightarrow 0$. That is, if we define $\delta t=t/n$ then $\Acal(t)=[\Acal(\delta t)]^n$ as $n\longrightarrow\infty$. Hence to show that a linear amplifier is phase covariant, all we have to do is show that its map satisfies \eqref{PhaseCovDefn} for an infinitesimal time interval $dt$. Since our counterexamples to the paramp conjecture in the main text are of the form $\Acal(t)=\exp(\Lcal t)$ with $\Lcal$ in the Lindblad form, we see that $\Acal(dt)=\mathbbm{1}+\Lcal\,dt$. The condition $\Acal(dt)\,\Phase=\Phase\,\Acal(dt)$ then becomes
\begin{align}
\label{InfPhasePres}
	\Lcal \, \Phase  = \Phase \, \Lcal   \;.
\end{align}
It is then possible to show that \eqref{InfPhasePres} is true. Here we will in fact prove that $\Acal(t)=\exp(\Lcal t)$ is a phase-covariant channel as long as $\Lcal$ is any $m$-photon dissipator. That is,
\begin{align}
\label{mPhotonCh}
	\Dcal[\ann^m] \, \Phase = \Phase \, \Dcal[\ann^m]  \;,   \quad    \Dcal[\adg{}^m] \, \Phase = \Phase \, \Dcal[\adg{}^m]  \;.
\end{align}
This covers both our counterexamples to the paramp conjecture. For $\Dcal[\ann^m]$ we have,
\begin{align}
	\Dcal[\ann^m] \, \Phase \, \rho = {}& \ann^m \, \rho_\varphi \, \adg{}^m - \frac{1}{2} \; \adg{}^m \, \ann^m \, \rho_\varphi
	                                      - \frac{1}{2} \; \rho_\varphi \, \adg{}^m \, \ann^m   \\[0.25cm]  
	= {}& \big( e^{-i\varphi\;\!\nhat} \,  e^{i\varphi\;\!\nhat} \big) \, \ann^m \, \rho_\varphi \, \adg{}^m \, \big( e^{-i\varphi\;\!\nhat} \,  e^{i\varphi\;\!\nhat} \big) 
	      - \frac{1}{2} \; \big( e^{-i\varphi\;\!\nhat} \, e^{i\varphi\;\!\nhat} \big) \, \adg{}^m \, 
	      \big( e^{-i\varphi\;\!\nhat} \, e^{i\varphi\;\!\nhat} \big) \, \ann^m \, \rho_\varphi  \nn \\
	    & - \frac{1}{2} \; \rho_\varphi \, \adg{}^m \,\big( e^{-i\varphi\;\!\nhat} \,  e^{i\varphi\;\!\nhat} \big)\, \ann^m \, 
	      \big( e^{-i\varphi\;\!\nhat} \,  e^{i\varphi\;\!\nhat} \big) \\[0.25cm]
\label{D[am]}
	= {}& e^{-i\varphi\;\!\nhat} \, \big( e^{i\varphi\;\!\nhat} \, \ann^m e^{-i\varphi\;\!\nhat} \big) \rho \, 
	      \big( e^{i\varphi\;\!\nhat} \, \adg{}^m e^{-i\varphi\;\!\nhat} \big) e^{i\varphi\;\!\nhat}  
	      - \frac{1}{2} \; e^{-i\varphi\;\!\nhat} \big( e^{i\varphi\;\!\nhat} \, \adg{}^m \, e^{-i\varphi\;\!\nhat} \big) 
	      \big( e^{i\varphi\;\!\nhat} \, \ann^m \, e^{-i\varphi\;\!\nhat} \big) 
	      \, \rho \, e^{i\varphi\;\!\nhat}  \nn \\
	    & - \frac{1}{2} \; e^{i\varphi\;\!\nhat} \, \rho \, \big( e^{i\varphi\;\!\nhat} \, \adg{}^m \, e^{-i\varphi\;\!\nhat} \big) 
	      \big( e^{i\varphi\;\!\nhat} \, \ann^m \, e^{-i\varphi\;\!\nhat} \big) e^{i\varphi\;\!\nhat}  \;.
\end{align}
This can simplified by noting that $\exp(i\varphi\;\!\nhat)\,\ann\,\exp(-i\varphi\;\!\nhat)=\ann\,\exp(-i\;\!\varphi)$ from which we can also see that
\begin{align}
\label{am}
	\exp(i\varphi\;\!\nhat) \, \ann^m \, \exp(-i\varphi\;\!\nhat) = \ann^m \, e^{-im\;\!\varphi}  \;.
\end{align}
Equation \eqref{D[am]} is thus
\begin{align}
	\Dcal[\ann^m] \, \Phase \, \rho = {}& e^{-i\varphi\;\!\nhat} \, \ann^m  \rho \, \adg{}^m \, e^{i\varphi\;\!\nhat}  
	                                      - \frac{1}{2} \; e^{-i\varphi\;\!\nhat} \, \adg{}^m \, \ann^m \, \rho \, e^{i\varphi\;\!\nhat}  
	                                      - \frac{1}{2} \; e^{i\varphi\;\!\nhat} \, \rho \, \adg{}^m \, \ann^m \, e^{i\varphi\;\!\nhat}  \\
	                                = {}& e^{-i\varphi\;\!\nhat} \, \bigg( \ann^m  \rho \, \adg{}^m \,   
	                                      - \frac{1}{2} \; \adg{}^m \, \ann^m \, \rho \,
	                                      - \frac{1}{2} \; \rho \, \adg{}^m \, \ann^m \, \bigg) \, e^{i\varphi\;\!\nhat}  \nn \\
\label{PhCovDam}     
	                                = {}& \Phase \, \Dcal[\ann^m] \, \rho  \;.
\end{align}
The proof for $\Dcal[\adg{}^m]$ follows similarly on replacing $\ann$ with $\adg$ and using 
\begin{align}
\label{adgm}
	\exp(i\varphi\;\!\nhat) \, \adg{}^m \, \exp(-i\varphi\;\!\nhat) = \adg{}^m \, e^{im\;\!\varphi}  \;.
\end{align}

\subsection{Phase sensitivity}

Aside from phase covariance, another important property of linear amplifiers is whether or not it is phase sensitive \cite{Cav82,SZ97}. This tries to capture whether the amplification and added noise due to the linear amplifier will differ for different directions in phase space. A linear amplifier is said to be phase insensitive if and only if for any value of $\varphi$, the quadrature
\begin{align}
	\hat{x}_\varphi = \frac{1}{\rt{2}} \; \big[ \ann \exp(-i\varphi) + \adg \exp(i\varphi) \big]  \; ,
\end{align}
 is such that it satisfies 
 \begin{gather}
 	\an{\hat{x}_\varphi(t)} = g \, \an{\hat{x}_\varphi(0)} \; ,    \\
	\an{[\Delta \hat{x}_\varphi(t)]^2} = g^2 \, \an{[\hat{x}_\varphi(0)]^2} + N  \; ,
\end{gather}
where $g$ and $N$ are independent of $\varphi$, and we have defined $\Delta\hat{x}_\varphi=\hat{x}_\varphi-\an{\hat{x}_\varphi}$. It is simple to show that for the noisiamp $g$ is as defined already, and 
\begin{align}
	N = g^2 \, \big( g^2 - 1 \big) \bigg[ \ban{\hat{n}(0)} +\frac{1}{2} \, \bigg] \; . 
\end{align}
It is clear from these relations that the noisiamp is phase insensitive.

\section{A three-photon counterexample}

To demonstrate the non-uniqueness of the noisiamp as an example which cannot be described by the paramp, we proposed in the main text the example defined by
\begin{align}
\label{3PhotonL}
	\ddt \; \rho(t) = \frac{\gamma}{9} \, \Dcal[\ann^3]\, \rho(t) + \frac{\gamma}{9} \, \Dcal[\adg{}^3] \, \rho(t) + \gamma \; \Dcal[\ann^2] \, \rho(t)  \;.
\end{align}
This is clearly a phase-covariant linear amplifier by the results of Sec.~\ref{PhaseCovarianceSM}. It is simple to show from this that
\begin{gather}
	\ddt \; \an{\ann} = \gamma \, \ban{\ann}  \;,  \\
\label{3Photon<n>}
	\ddt \; \an{\nhat} = 2\,\gamma\,\ban{\nhat^2} + 6\,\gamma\,\ban{\nhat} + 2 \, \gamma   \;.  
\end{gather}
It is obvious that $\an{\ann(t)}=g\an{\ann(0)}$ where $g=\exp(\gamma\,t)$.
\begin{figure}[t]
\centerline{\includegraphics[width=0.6\textwidth]{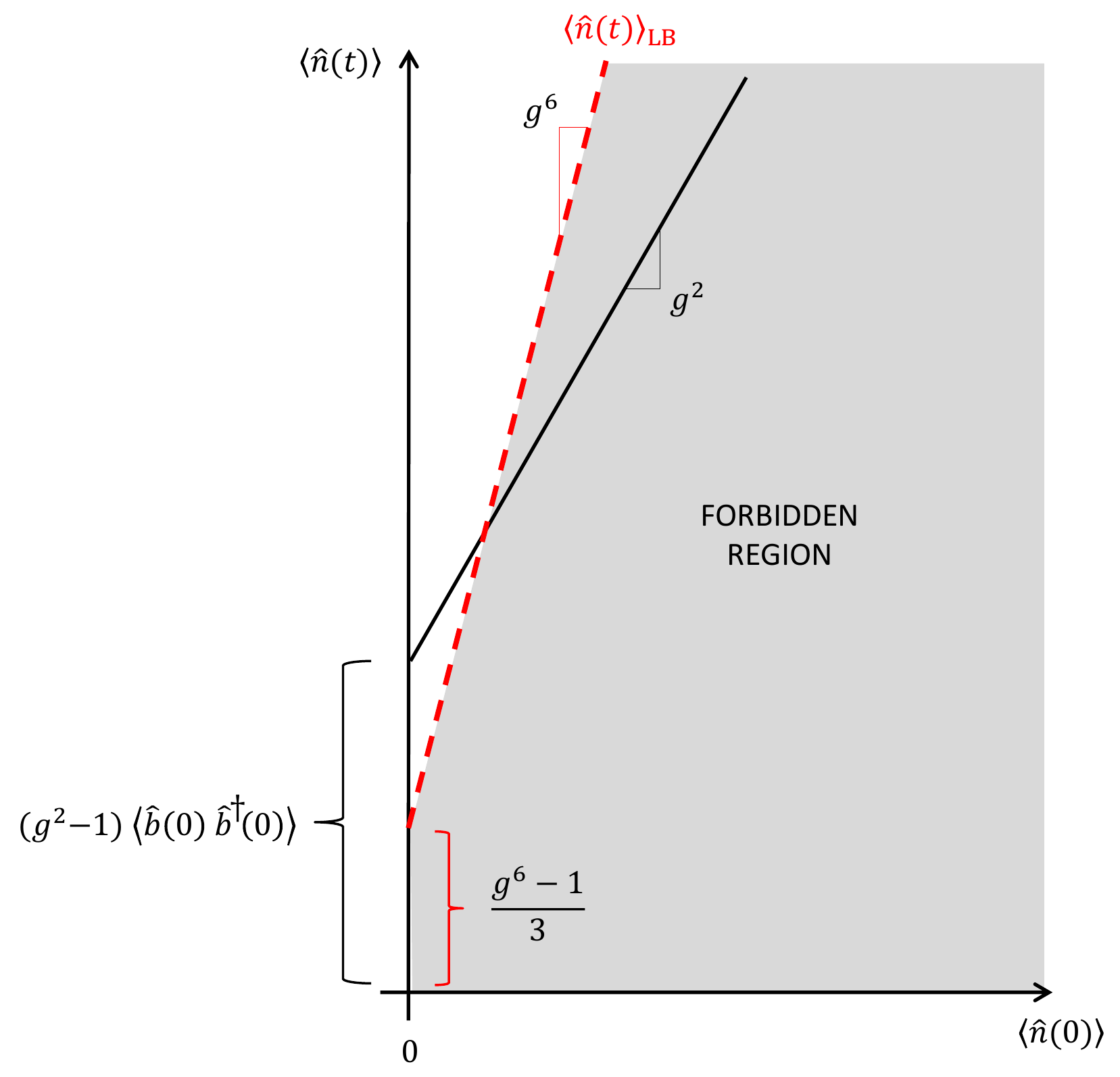}}
\caption{\label{ThreePhotonEx} No matter how the paramp operating state $\sigma$ is chosen to change $\an{\hat{b}(0)\,\hat{b}\dg(0)}$, it cannot produce photon-number outputs that are entirely within the allowed region above the red-dashed line.}
\end{figure}
However, the output average photon number is now coupled to its second moment. We can still show that it leads to unattainable values for the paramp by considering a lower bound of $\an{\nhat(t)}$ by ignoring the first term of \eqref{3Photon<n>}. Solving the resulting differential equation gives
\begin{align}
\label{<n>LB}
	\ban{\nhat(t)}_{\rm LB} = {}& g^6 \, \ban{\nhat(0)} + \frac{g^6-1}{3}  \le  \ban{\hat{n}(t)}  \; .
\end{align}
The paramp with identical amplitude gain has
\begin{align}
\label{<n>ParametricSM}
	\ban{\hat{n}(t)} = {}& g^2 \ban{\hat{n}(0)} + \big(g^2-1\big)\,\ban{\hat{b}(0)\,\hat{b}\dg(0)}  \;.
\end{align}
When considered a function of $\an{\hat{n}(0)}$, \eqref{<n>ParametricSM} is a straight line with gradient $g^2$ and vertical intercept $(g^2-1)\,\an{\hat{b}(0)\,\hat{b}\dg(0)}$. This is shown as the black line in Fig.~\ref{ThreePhotonEx}. On the same axes \eqref{<n>LB} is shown as the red dashed line (not to scale). It is also a straight line but with a larger gradient and vertical intercept $(g^6-1)/3$. The actual solution to \eqref{3Photon<n>} must therefore lie above the red-dashed line while the area below it (shaded region) is forbidden. Figure~\ref{ThreePhotonEx} clearly illustrates that no matter how $\an{\hat{b}(0)\,\hat{b}\dg(0)}$ is chosen (by choosing $\sigma$ in the ancillary mode), the paramp $\an{\nhat(t)}$ always has a segment in the forbidden region of the three-photon example.


\begin{thebibliography}{100}


\bibitem{CTD+80}
C. M. Caves, K. S. Thorne, R. W. P. Drever, V. D. Sandberg, and M. Zimmermann, Rev. Mod. Phys. {\bf 52}, 341 (1980).
	
	
\bibitem{CDG+10}
A. A. Clerk, M. H. Devoret, S. M. Girvin, F. Marquadt and R. J. Schoelkopf, Rev. Mod. Phys. {\bf 82}, 1155 (2010).


\bibitem{ZFB10}
A. Zavatta, J. Fiur\'{a}\v{s}ek, and M. Bellini, Nat. Photon. {\bf 5}, 52 (2010).

\bibitem{VMS+12}
R. Vijay, C. Macklin, D. H. Slichter, S. J. Weber, K. W. Murch, R. Naik, A. N. Korotkov, and I. Siddiqi, Nature {\bf 490}, 77 (2012)


\bibitem{HKL+14}
F. Hudelist, J. Kong, C. Liu, J. Jing, Z. Y. Ou, and W. Zhang, Nat. Comm. {\bf 5}, 3049 (2014).


\bibitem{RL09}
T. C. Ralph and A. P. Lund, \emph{Proceedings of the 9th International Conference on Quantum Communication Measurement and Computing, (A. Lvovsky Ed.)}, 155, (AIP, 2009).



\bibitem{XRLWP10}
G. Y. Xiang, T. C. Ralph and N. Walk and G. J. Pryde, Nat. Photon. {\bf 4}, 316 (2010).


\bibitem{CWA+14}
H. M. Chrzanowski, N. Walk, S. M. Assad, J. Janousek, S. Hosseini, T. C. Ralph, T. Symul, and P. K. Lam, Nat. Photon. {\bf 8}, 333 (2014).


\bibitem{HZD+16}
J. Y. Haw, J. Zhao, J. Dias, S. M. Assad, M. Bradshaw, R. Blandino, T. Symul, T. C. Ralph, and P. K. Lam, Nat. Comm. {\bf 7}, 1 (2016).


\bibitem{PvE19}
T. B. Propp and S. J. van Enk, Opt. Express {\bf 27}, 23454 (2019).


\bibitem{Cav82maintext}
C. M. Caves, Phys. Rev. D {\bf 26}, 1817 (1982).


\bibitem{CCJP12}
C. M. Caves, J. Combes, Z. Jiang, and S. Pandey, Phys. Rev. A {\bf 86}, 063802 (2012).


\bibitem{CHN+19}
See Supplementary Material. 


\bibitem{Lin76}
G. Lindblad, Comm. Math. Phys. {\bf 48}, 119 (1976).


\bibitem{BP02}
H.-P. Breuer and F. Petruccione, \emph{The Theory of Open Quantum Systems}, (Oxford University Press, 2002).


\bibitem{Car02}
H. J. Carmichael, \emph{Statistical Methods in Quantum Optics 1} (Second-corrected-printing), (Springer, 2002).


\bibitem{DH14}
P. D. Drummond and M. Hillery, \emph{The Quantum Theory of Nonlinear Optics}, (Cambridge University Press, 2014)


\bibitem{Aga13}
G. S. Agarwal, \emph{Quantum Optics}, (Cambridge University Press, 2013).


\bibitem{SZ97}
M. O. Scully and M. S. Zubairy, \emph{Quantum Optics} (Cambridge University Press, 1997).


\bibitem{Lam67}
P. Lambropoulos, Phys. Rev. {\bf 156}, 286 (1967).


\bibitem{MW74}
K. J. McNeil and D. F. Walls, J. Phys. A {\bf 7}, 617 (1974).


\bibitem{HNGO11}
A. Hayat, A. Nevet, P. Ginzburg, and M. Orenstein, Semicond. Sci. Technol. {\bf 26}, 083001 (2011).


\bibitem{Gau03}
D. J. Gauthier, Prog. Opt. {\bf 45}, 205 (2003).


\bibitem{NEFE77}
L. M. Narducci, W. W. Edison, P. Furcinitti, and D. C. Eteson, Phys. Rev. A {\bf 16}, 1665 (1977).


\bibitem{NZT81}
B. Nikolaus, D. Z. Zhang, and P. E. Toschek, Phys. Rev. Lett. {\bf 47}, 171 (1981).


\bibitem{BRGDH87}
M. Brune, J. M. Raimond, P. Goy, L. Davidovich, and S. Haroche, Phys. Rev. Lett. {\bf 59}, 1899 (1987).


\bibitem{AGBZ90}
I. Asharaf, J. Gea-Banacloche, and M. S. Zubairy, Phys. Rev. A {\bf 42}, 6704 (1990).


\bibitem{Iro92}
C. N. Ironside, IEEE J. Quantum Electron. {\bf 28}, 842 (1992).


\bibitem{GWMM92}
D. J. Gauthier, Q. Wu, S. E. Morin, and T. W. Mossberg, Phys. Rev. Lett. {\bf 68}, 464 (1992).


\bibitem{NHO10}
A. Nevet, A. Hayat, and M. Ornstein, Phys. Rev. Lett. {\bf 104}, 207404 (2010).


\bibitem{RSSHS16}
M. Reichert, A. L. Smirl, G. Salamo, D. J. Hagan, and E. W. Van Stryland, Phys. Rev. Lett. {\bf 117}, 073602 (2016).


\bibitem{MRBB18}
S. Melzer, C. Ruppert, A. D. Bristow, and M. Betz, Opt. Lett. {\bf 43}, 5066 (2018).


\bibitem{GZ10}
C. W. Gardiner and P. Zoller, \emph{Quantum Noise} (Third edition), (Springer, 2010).


\bibitem{Ito42}
K. \ito, J. Pan-Japan Math. Coll. {\bf 1077}, 1352 (1942).


\bibitem{Ito44}
K. \ito, Proc. Imp. Acad. Tokyo {\bf 20}, 519 (1944).


\bibitem{Ito46}
K. \ito, Proc. Imp. Acad. Tokyo {\bf 22}, 32 (1946).


\bibitem{HP84}
R. L. Hudson and K. R. Parthasarathy, Commun. Math. Phys. {\bf 93}, 301 (1984).


\bibitem{WM10}
H. M. Wiseman and G. J. Milburn, \emph{Quantum Measurement and Control}, (Cambridge University Press, 2010).


\bibitem{Mil19}
P. W. Milonni, \emph{An Introduction to Quantum Optics and Quantum Fluctuations} (Oxford University Press, 2019).


\bibitem{Chi15}
M.-H. Chiang, \emph{Quantum Stochastics}, (Cambridge University Press, 2015).


\bibitem{Par92}
K. R. Parthasarathy, \emph{An Introduction to Quantum Stochastic Calculus}, (Birkh\"{a}user, 1992).


\bibitem{WZ65}
E. Wong and M. Zakai, Int. J. Engng. Sci. {\bf 3}, 213 (1965).


\bibitem{Gou06}
J. Gough, J. Math. Phys. {\bf 47}, 113509 (2006).


\bibitem{GC85}
C. W. Gardiner and M. J. Collett, Phys. Rev. A {\bf 31}, 3761 (1985).


\bibitem{Jac10}
K. Jacobs, \emph{Stochastic Processes for Physicists: Understanding Noisy Systems}, (Cambridge University Press, 2010).


\bibitem{Gar09}
C. Gardiner, \emph{Stochastic Methods} (Fourth edition), (Springer, 2009).


\bibitem{HGO08}
A. Hayat, P. Ginzburg, and M. Orenstein, Nat. Photonics {\bf 2}, 238 (2008).


\bibitem{NBH+10}
A. Nevet, N. Berkovitch, A. Hayat, P. Ginzburg, S. Ginzach, O. Sorias, and M. Orenstein, NanoLett {\bf 10}, 1848 (2010).
 

\bibitem{OIKA11}
Y. Ota, S. Iwamoto, N. Kumagai, and Y. Arakawa, Phys. Rev. Lett. {\bf 107}, 233602 (2011).


\bibitem{SMrefs}
See Supplementary Material [url] for how an ion-trap realisation may be accomplished which includes Refs.~\cite{LBMW03,LS13}


\bibitem{LBMW03}
D. Leibfried, R. Blatt, C. Monroe, and D. Wineland, Rev. Mod. Phys. {\bf 75}, 281 (2003).


\bibitem{LS13}
T. E. Lee and H. R. Sadeghpour, Phys. Rev. Lett. {\bf 111}, 234101 (2013).



%
%
%
%




%
%
%
%
%
%
%
%
%
%
%
%
%
%
%
%
%
%









%
%
%
%
%
%
%
%
%
%
%
%
%
%
%
%
%
%
%
%
%
%
%
%
%
%
%
%
%
%
%
%
%
%
%
%
%
%
%
%
%
%
%





\end{thebibliography}

\begin{thebibliography}{100}


\bibitem{DH14}
P. D. Drummond and M. Hillery, \emph{The Quantum Theory of Nonlinear Optics}, (Cambridge University Press, 2014).


\bibitem{BP02SM}
H.-P. Breuer and F. Petruccione, \emph{The Theory of Open Quantum Systems}, (Oxford University Press, 2002).


\bibitem{Car02SM}
H. J. Carmichael, \emph{Statistical Methods in Quantum Optics 1} (Second-corrected-printing), (Springer, 2002).


\bibitem{Gar09}
C. Gardiner, \emph{Stochastic Methods} (Fourth edition), (Springer 2009).


\bibitem{Jac10a}
K. Jacobs, \emph{Stochastic Processes for Physicists: Understanding Noisy Systems}, (Cambridge University Press, 2010).


\bibitem{Par92}
K. R. Parthasarathy, \emph{An Introduction to Quantum Stochastic Calculus}, (Birkh\"{a}user, 1992).


\bibitem{GZ10SM}
C. W. Gardiner and P. Zoller, \emph{Quantum Noise} (Third edition), (Springer, 2010).


\bibitem{WZ65SM}
E. Wong and M. Zakai, Int. J. Engng. Sci. {\bf 3}, 213 (1965).


\bibitem{Gou06SM}
J. Gough, J. Math. Phys. {\bf 47}, 113509 (2006).


\bibitem{GC85}
C. W. Gardiner and M. J. Collett, Phys. Rev. A {\bf 31}, 3761 (1985).


\bibitem{leibfried2003quantum} 
D. Leibfried, R. Blatt, C. Monroe, and D. Wineland, Rev. Mod. Phys. {\bf 75}, 281 (2003).
	
	
\bibitem{lee2013quantum} 
T. E. Lee and H. R. Sadeghpour, Phys. Rev. Lett. {\bf 111}, 234101 (2013).


\bibitem{WM10SM}
H. M. Wiseman and G. J. Milburn,  \emph{Quantum Measurement and Control}, (Cambridge University Press, 2010).


\bibitem{Chi15}
M.-H. Chiang, \emph{Quantum Stochastics}, (Cambridge University Press, 2015).


\bibitem{CCJP12SM}
C. M. Caves, J. Combes, Z. Jiang, and S. Pandey, Phys. Rev. A {\bf 86}, 063802 (2012).


\bibitem{Cav82}
C. M. Caves, Phys. Rev. D {\bf 26}, 1817 (1982). 


\bibitem{SZ97}
M. O. Scully and M. S. Zubairy, \emph{Quantum Optics} (Cambridge University Press, 1997).



\end{thebibliography}
\end{document}